\documentclass[11pt,english,review,authoryear]{elsarticle}
\usepackage[T1]{fontenc}
\usepackage[latin9]{inputenc}
\usepackage{geometry}
\usepackage{amsthm}
\usepackage{amsmath}
\usepackage{graphicx}
\usepackage{setspace}
\usepackage[authoryear]{natbib}
\makeatletter
\numberwithin{equation}{section}
\numberwithin{figure}{section}
\theoremstyle{plain}
\newtheorem{thm}{\protect\theoremname}
  \theoremstyle{remark}
  \newtheorem{claim}[thm]{\protect\claimname}
  \theoremstyle{plain}
  \newtheorem{lem}[thm]{\protect\lemmaname}

\journal{Games and Economic Behavior}

\makeatother

\usepackage{babel}
  \providecommand{\claimname}{Claim}
  \providecommand{\lemmaname}{Lemma}
\providecommand{\theoremname}{Theorem}

\begin{document}

\title{Group Formation with a Network Constraint}

\author[rvt]{Katharine A. Anderson\fnref{fn1}\corref{cor1}}

\ead{andersok@andrew.cmu.edu}

\fntext[fn1]{Carnegie Mellon University, Tepper School of Business}

\cortext[cor1]{Corresponding author}

\address[rvt]{Carnegie Mellon University, Tepper School of Business, 5000 Forbes
Ave, Pittsburgh, PA 15213}
\begin{abstract}
Group formation is important in many economic contexts. The current
literature on group formation assumes that individuals may join any
existing group. In this paper, I consider the implications of social,
geographic, and informational constraints to group membership decisions.
I embed the players in a network of relationships, which constrains
their choice of groups--they may only join a group if that group contains
a member that they are connected to on the network. I then examine
how this network constraint affects the equilibrium group structure.
I show that even with complete information, unconstrained individuals
form groups that are inefficiently large. When individuals are constrained,
the resulting group structures are much closer to the socially optimal
group structure, because the constraint limits the ability of the
individual to free ride on the efforts of other group members. The
efficiency of the outcome is related to the structure of the network
constraint--outcomes are more efficient when networks are sparse and
have few random connections. \end{abstract}
\maketitle
Group formation is important in a variety of economic contexts: within
firms, workers form teams to combine human capital resources and aid
in production; farmers form cooperatives to pool outcomes and share
risk; consumers create groups to increase buying power, and rent-seekers
form coalitions to increase their leverage. In a social context, groups
are created to organize volunteer efforts, lobby for political causes,
and create social unity. In all of these contexts, there is a benefit
to the individual from being a member of the group, and the success
of both the individual and the overall society depends on group structure.
The current literature on group formation assumes that individuals
are unconstrained in their choice of groups. In other words, any individual
can join any group at any time. However, instances where group membership
decisions are truly unconstrained are rare--individuals face a wide
range of social, geographic, and informational constraints in their
group membership decisions. In this paper, I extend the existing group
membership models to account for such widespread constraints. I embed
individuals in a network of relationships, which constrains their
choice of groups--an individual may only join a group which contains
one of her contacts on the network. For example, an individual might
be constrained to joining groups containing a friend or relative.
I then examine how this network constraint affects the equilibrium
group structure. I show that in some cases, network constraints make
equilibrium group structures more efficient, and result in higher
social welfare. I also show that we can see hints of the underlying
network structure in the structure of the groups that form. 

The literature on group formation spans a number of subfields, including
industrial organization, political economy, and public economics.%
\footnote{This diversity of fields induces a diversity of terminology. Alternatives
to {}``group formation'' include \textquotedbl{}club formation,\textquotedbl{}
\textquotedbl{}team assembly,\textquotedbl{} and \textquotedbl{}coalition
formation.\textquotedbl{}%
} All games in this literature have three common elements: 1) players
make decisions about group membership, 2) they can be a member of
one and only one group, 3) their payoffs depend on the arrangement
of the players into groups. In the earliest work, players make their
group membership decisions simultaneously.%
\footnote{See, for example, \citet{Hart1983}, \citet{Nitzan1991}, \citet{Yi2000},
\citet{Konishi1997}, and \citet{Heintzelman2009}.%
} However, more recent literature has focused on dynamic group formation
games, in which players make their group formation decisions sequentially
over time.%
\footnote{See \citet{Demange2005} and \citet{Bloch2010} for a survey of this
work. This includes, among others, \citet{Bloch1996}, \citet{Yi2000},
\citet{Arnold2002}, \citet{Konishi2003}, \citet{Arnold2005}, \citet{Macho-Stadler2006},
and \citet{Page2007a}.%
} 

This paper further extends this literature by considering the very
real constraints that individuals face when making group membership
decisions. Consider, for example, a set of farmers forming water management
groups along the banks of a river. Although it is conceivable that
the farmers would organize into groups at random, they are more likely
to organize with farmers who are adjacent to them on the river than
those in distant locations. Other groups are governed by social connections.
For example, research lab groups are more likely to be composed of
colleagues than strangers. In some cases, social constraints on group
membership even serve a purpose by facilitating enforcement of rules
and social norms. There may also be informational constraints to group
membership--an individual can only join an organization if she knows
of its existence. Covert organizations are an extreme example of informational
constraints. 

A network is a natural way of representing constraints on group membership.
Individuals are embedded in a fixed, exogenous network of relationships,
which limit their access to other groups, and an individual can only
join a group if she is connected to a current member on the network.
This method allows me to use machinery from the burgeoning networks
literature, which explores how network structure affects individual
behavior.%
\footnote{See \citet{Jackson2008} for a survey of the ways that limiting interactions
between individuals can affect strategic behavior. \citet{Girvan2002},
\citet{Newman2003}, and \citet{Copic2009} look at methods for finding
community structure in social networks without overt group membership.%
} Depending on its structure, this network may represent any of the
constraints mentioned above. Networks representing spacial constraints
will have fewer random connections than the networks representing
social constraints. The density of the network constraint will reflect
how close the ties have to be in order to allow group membership.
For example, becoming a member of a volunteer organization may only
require a passing relationship with a current member, whereas an individual
joining a covert organization will likely only find acceptance if
she is extremely close to a current member. In other words, if we
imagine that there is a threshold level of familiarity that is required
for group membership, that level of familiarity will dictate the density
of the network. A lower threshold of familiarity implies more links,
which implies a denser network.

This network structure allows us to consider whether different types
of constraints create different types of group structures. Do individuals
behave differently when their group membership decisions are moderated
by spacial constraints, as opposed to social constraints? How does
equilibrium group structure change as the requirements for membership
become more stringent? In other words, can we see traces of network
structure in the structure of groups?

I start with a game in which individuals are completely unconstrained
in their choice of groups. I show that when individuals are unconstrained
in their group membership, they will form groups that are much too
large, from the standpoint of social welfare. This is a somewhat surprising
result, which has not yet been noted in the group formation literature.
Groups become too large because of the externality that new members
impose on the group's existing membership--new members free ride off
of the efforts of early group members, and new groups are under-provided.
I then consider the effect of a network constraint on equilibrium
group structure. I show that when individuals are constrained by their
networks, the resulting group structures are much closer to the socially
optimal group structure, and thus much more efficient. This is because
the network constraint limits the ability of the individual to free
ride on the efforts of other group members. The efficiency of the
outcome is related to the structure of the network constraint--outcomes
are more efficient when networks are sparse. This suggests that group
structure will be more efficient when the constraints on group membership
are more severe. Moreover, I show that holding the density of the
network constant, outcomes are more efficient in networks with fewer
random connections. This suggests a secondary effect--local constraints
are more binding than social constraints, making the resulting group
structure more efficient.

\section{Dynamic Group Formation Game}

Let $I=\left\{ 1,2,...N\right\} $ be a set of $N$ homogeneous individuals.
An individual can be a member of one and only one group--thus, the
group structure at time $t$ is a partition of $I$, $\pi(t)=\left\{ G_{1}G_{2}...G_{J(t)}\right\} $,
where $G_{j}$ denotes the set of individuals in group $j$.%
\footnote{Note that the set of individuals in a group is also a function of
time. That is, $\pi\left(t\right)=\left\{ G_{1}\left(t\right),G_{2}\left(t\right),...,G_{J\left(t\right)}\left(t\right)\right\} $.
However, for notational clarity, I will suppress the time-dependance.%
} Note that the number of groups is determined endogenously, and thus
$J(t)$ may vary from one period to the next. The set of all such
partitions of the players into groups is denoted $\Pi$.

The players have identical payoff functions that depend on the size
of the player's own group: $f(g_{j})$ where $i\in G_{j}$ and $g_{j}=\left|G_{j}\right|$
is the size of group $j$.%
\footnote{The assumption that payoffs depend only on own group size obviously
does not allow for externalities between groups, nor does it allow
players to have preferences over group composition. However, this
is an appropriately simple starting point for dynamic analysis--to
the extent that inter-coalition externalities muddy behavior, they
are best left to future extensions.%
} I assume that $f(g)$ is single-peaked with maximum value $g^{*}$.%
\footnote{The single-peak assumption is useful because individual and social
preferences are aligned (the individuals all want to be in groups
of size $g^{*}$, and social welfare is highest when this occurs)
and as I will show, the equilibrium reached is suboptimal, \emph{despite
this alignment}. This assumption is violated if there are several
group sizes that are local maxima--for example a 4th order polynomial
will sometimes have two peaks in the positive range (eg: $-x^{4}+15x^{3}-74x^{2}+132x-38$)
. However, it is relatively easy to extend the results here to functions
with two or even more peaks, so long as the function does not become
too noisy. You simply think of each peak as an individual single-peak
function. %
}

Since individuals in this game are homogeneous, the exact arrangement
of the players in the groups is not as important as the sizes of the
groups. Thus, I will often find it convenient to refer to the vector
of group sizes resulting from a particular partition of the individuals,
rather than referring to the partition itself: define the \emph{group
size vector} of a partition $\pi(t)=\left\{ G_{1}...G_{J}\right\} $
by $\left\langle g_{1}...g_{J}\right\rangle $. 

Players move sequentially in an order of motion, $\phi$. At time
$t$, the active player can choose to either join an existing group,
$G_{j}\in\pi(t)$, or strike out on her own, forming a group of size
1. Thus, individual $i$'s action set at time $t$ can be denoted
by $A_{i}(t)=\pi(t)\cup\emptyset$, where $\emptyset$ denotes the
action of striking out as an individual. Thus, $\left(N,f(g),\phi\right)$
defines a dynamic group formation game. 

I will assume that players make their group membership decisions myopically--that
is, they decide which group will maximize their return, given only
the \emph{current} group structure. This defines a behavior strategy,
$\beta\left(.\right)$, which maps the current group partition, $\pi(t)$,
to the individual's action set: $\beta\left(\pi(t)\right)\in A_{i}(t)$.
This myopic behavior strategy is similar to that used in the sequential
group formation literature, including \citet{Arnold2002} and \citet{Arnold2005}.
Myopia is behaviorally quite realistic in this context--even with
small numbers of players, the calculations required of a far-sighted
player quickly become unreasonable. Moreover, as we will see in later
sections, the cognitive capacity constraint binds even more heavily
when a network constraint is considered.%
\footnote{Note that it is possible to relax the myopia assumption. In particular,
if players discount the future, there exists a discount rate, $\delta\in\left(0,1\right)$,
such that the main result of this section (Theorem \ref{thm:number seq eq})
still holds. This discounting could represent either a traditional
discounting of future payoffs, or the cognitive limitations of the
player. %
}

The outcome of a myopic Nash equilibrium of this dynamic game is a
partition of the players into groups, $\pi^{*}=\left\{ G_{1}...G_{J}\right\} $
such that $f\left(g_{j}\right)\ge f\left(g_{k}+1\right)\forall G_{j},G_{k}\in\pi^{*}$.%
\footnote{In other words, a group structure is an equilibrium if it is stable
against unilateral, myopic deviations. It is worth noting that this
equilibrium concept differs from that used in \citet{Arnold2005}.
They consider a {}``Nash Club Equilibrium'' (a group structure which
is stable to deviations by coalitions of individuals within a particular
group) and a {}``k-remainder Nash Club Equilibrium'' (which is stable
to deviations when k individuals are dropped from the system). I have
used the myopic Nash Equilibrium because it is simpler. It is worth
noting that I obtain dramatically different results using this equilibrium
concept than Arnold and Wooders do using the Nash Club and k-remainder
Equilibria.%
} Let $\varepsilon\left(N,f(g),\phi\right)$ denote the set of equilibrium
group size vectors for the game $\left(N,f\left(g\right),\phi\right)$.

\begin{figure}[h]
\includegraphics[scale=0.6]{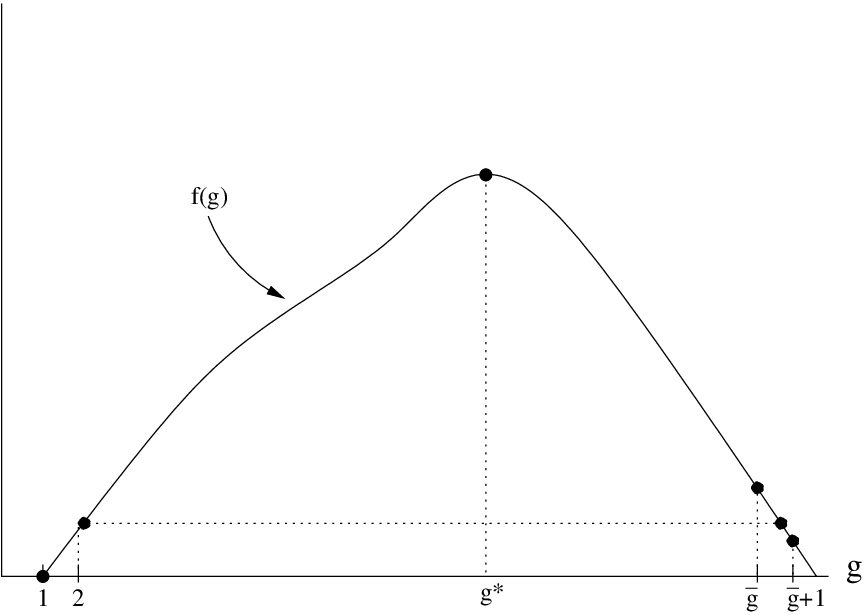}

\caption{\label{Illustrating gbar}An illustration of $\bar{g}$ --the smallest
$g$ such that $f(g+1)<f(2)$. }
\end{figure}

Finally, it will be convenient to define one additional feature of
the utility function: define $\bar{g}$ to be the smallest $g$ such
that $f(g+1)<f(2)$. Note that $\bar{g}$ is the largest group that
will form before an individual forms a new group of size 2.%
\footnote{If $f(N)>f(2)$, then no one will ever find it profitable to form
a new group of size 2. For convenience, I will define $\bar{g}=N$
in these cases.%
} Figure \ref{Illustrating gbar} illustrates an example of $\bar{g}$.

\subsection{Stable Group Configurations}

The equilibrium coalition group partition of the dynamic group formation
game must necessarily be stable group configurations--partitions of
the players into groups such that no individual wishes to deviate
unilaterally. These stable group configurations are Nash equilibria
of the static group formation game, where $N$ players make their
group membership decisions simultaneously, based on their payoff $f\left(g\right)$.
Thus, before considering the equilibrium of the dynamic game, it is
useful to consider the equilibrium of this static game. 

A static group formation game is defined by $\left(N,f\left(g\right)\right)$.
Call the set of Nash equilibrium group size vectors for this static
game $\epsilon\left(N,f\left(g\right)\right)$. This is the set of
stable group configurations. Theorem \ref{thm:static eq} characterizes
$\varepsilon\left(N,f(g)\right)$, and thus the set of stable group
configurations. This theorem highlights two characteristics of any
stable group configuration: 1) the groups will mostly be larger than
the social optimum (at most one will be smaller) and 2) all of the
groups larger than the optimum will be approximately the same size. 
\begin{thm}
\label{thm:static eq}Let $\left(N,f(g)\right)$ be a static group
formation game with single-peaked payoff function $f(g)$. The set
of Nash Equilibria of that game, $\varepsilon\left(N,f(g)\right)$,
is the union of two sets: \end{thm}
\begin{enumerate}
\item $\left\{ \left\langle g_{1}...g_{J}\right\rangle |\bar{g}\ge g_{j}\ge g^{*}\,\forall j\, and\,|g_{j}-g_{k}|\le1\,\forall j,k\right\} $ 
\item $\left\{ \left\langle g_{1}...g_{J}\right\rangle |g_{1}<g*,\, g_{j}\ge g^{*}\,\forall j\ne1,\, g_{j}=g_{k}\,\forall j,k\ne1,\right.$\\
 $\left.and\, f(g_{k})\ge f(g_{1}+1)\ge f(g_{1})\ge f(g_{k}+1)\right\} $ \end{enumerate}
\begin{proof}
See Appendix.
\end{proof}

\subsection{Inefficiency of Equilibria in the Dynamic Game}

One insight gained from the static game is that that there will often
be multiple stable group configurations (see Appendix for further
information). This raises the question: in a dynamic environment,
which of these stable configurations will be an equilibrium? And will
any of those equilibria be socially optimal? Theorem \ref{thm:number seq eq}
states that when players start the game as individuals,%
\footnote{Obviously the equilibrium reached will depend on the initial condition.
Starting the game with the individuals acting alone seems very natural.
The results that follow are unchanged if the individuals start the
game in a grand coalition.%
} a dynamic group formation game has a unique equilibrium group size
vector, $\gamma\left(N,f(g)\right)=\left\langle g_{1}...g_{J}\right\rangle $,%
\footnote{Note that the mapping from partitions to group size vectors is many-to-one,
and thus the mapping from equilibrium partitions to equilibrium group
size vectors will be as well.%
} which does not depend on the order of motion, $\phi$. Moreover,
this equilibrium is always the worst possible stable group configuration
from the standpoint of social welfare, despite the alignment between
social and individual preferences implied by single-peaked utility.
\begin{thm}
\label{thm:number seq eq}Let $\left(N,f(g),\phi\right)$ be a dynamic
group formation game, with $f(g)$ single-peaked, and $\pi\left(0\right)=\left\{ \left\{ 1\right\} ,\left\{ 2\right\} ,...,\left\{ N\right\} \right\} $.
Then there is a unique myopic Nash equilibrium group size vector,
$\gamma\left(N,f(g)\right)=\left\langle g_{1}...g_{J}\right\rangle $,
which is not a function of the order of motion, $\phi$. Moreover,
$\gamma\left(N,f\left(g\right)\right)=\arg\min_{\varepsilon\left(N,f(g)\right)}\sum_{i\in I}f(g_{i})$.
That is, the myopic Nash Equilibrium outcome of the dynamic game is
the stable configuration that minimizes social welfare. \end{thm}
\begin{proof}
Let $\left(N,f(g),\phi\right)$ be an arbitrary sequential group formation
game. Any equilibrium of $\left(N,f(g),\phi\right)$ must, necessarily,
be an equilibrium of the corresponding static game, $\left(N,f(g)\right)$.
I will show that regardless of the order of motion, $\phi$, the players
will settle into the configuration, $\left\langle g_{1}...g_{J}\right\rangle \in$
$\epsilon\left(N,f\left(g\right)\right)$ where groups are the largest.
This configuration yields the lowest social welfare of all possible
stable configurations. 

First, note that the stable configuration with the lowest possible
social welfare is the one with the smallest number of groups--$\frac{N-\bar{r}}{\bar{g}}+1$
total groups, where $\bar{r}=Nmod\bar{g}>0$.%
\footnote{Or $\frac{N}{\bar{g}}$ groups if $Nmod\bar{g}=0$.%
} If $f(g)$ is strictly increasing or decreasing, then the groups
will wind up in this configuration trivially. So suppose $f(g)$ is
unimodal. $f(g)$ unimodal implies $f(1)<f(2)$. Thus, regardless
of the order of motion, the first individual will always want to start
a new group. Since $f(g+1)>f(2)\,\forall\, g<\bar{g}$, all subsequent
individuals will prefer to join the existing group to forming a new
group of size 2. It is only worthwhile to create a second group of
size 2 when the existing group is size $\bar{g}$. More generally,
it will only be worth forming a new group of size 2 when all existing
groups have reached size $\bar{g}$. Thus, the final group forms when
there are $\frac{N-\bar{r}}{\bar{g}}$ groups of size $\bar{g}$,
creating a total of $\frac{N-\bar{r}}{\bar{g}}$ groups of size $\bar{g}$
and one smaller group. The groups will change size in subsequent turns.
However, regardless of the order of motion, no individual will ever
choose to form a group of size 1, because $f\left(1\right)<f\left(2\right)<f\left(\bar{g}\right)$.
Thus, the equilibrium reached is one with $\frac{N-\bar{r}}{\bar{g}}+1$
groups of the largest possible size, and thus the lowest possible
social welfare. %
\footnote{Note that this result is due, in part, to the fact that the myopic
Nash Equilibrium considers only unilateral deviations. If we allow
a subgroup of up to $n$ individuals to make their membership decisions
as a group, then any equilibrium that exists will necessarily have
smaller groups. However, the set of equilibria that are stable to
such coalitional deviations are largely empty (see \citet{Arnold2005}).
More importantly, when we move on to games with a network constraint,
as in the following section, it becomes less clear what is meant by
a configuration that is stable to {}``coalitional deviations.''
Analysis of more complicated, network-specific coalitional equilibrium
concepts are obviously venues for future work.%
}
\end{proof}

\subsection{An Example with Logistic Utility}

\begin{figure}
\begin{description}
\item [{\includegraphics[scale=0.65]{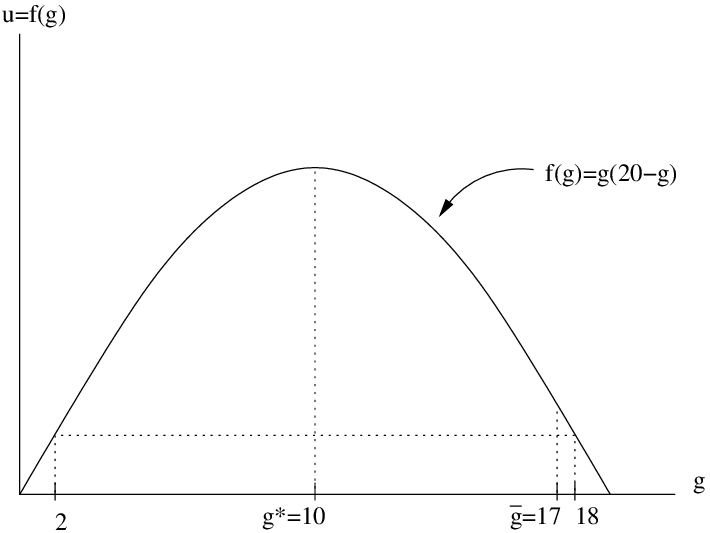}}]~
\end{description}
\caption{\label{fig:Logistic Function}Logistic payoff function: $f\left(g\right)=g\left(20-g\right)$.
The highest payoff in a group of size $g^{*}=10$.}

\end{figure}

This result can best be understood via an example. Consider a dynamic
group formation game with 100 players and a logistic payoff function
$f(g)=g(20-g)$. This function is single-peaked with maximum $g^{*}=10$
and $\bar{g}=17$. It is illustrated in Figure \ref{fig:Logistic Function}.

First, consider the possible stable group configurations. In any stable
group configuration, at most one coalition will be smaller than the
socially optimal group size, $g^{*}=10$. Moreover, all of the groups
larger than the social optimum will be approximately the same size.
Using these two facts, one can show that there are 5 stable group
configurations with group size vectors $\left\langle 10,10,10,10,10,10,10,10,10,10\right\rangle $,
$\left\langle 11,11,11,11,11,11,11,11,12\right\rangle $, $\left\langle 12,12,12,12,13,13,13,13\right\rangle $,
$\left\langle 14,14,14,14,14,15,15\right\rangle $, and $\left\langle 16,16,17,17,17,17\right\rangle $.
Note that $\left\langle 20,20,20,20,20\right\rangle $ is \emph{not}
a stable configuration, because an individual in a group of size 20
is better off striking out as an individual. Note that only one of
these stable configurations is efficient: the one where all groups
are size 10. This is, not coincidentally, the stable configuration
with the smallest possible group sizes. This will, in fact, always
be the case, regardless of the utility function.

Now, consider the dynamic group formation process. Theorem \ref{thm:number seq eq}
indicates that there will be a unique equilibrium, and moreover, that
equilibrium will be the stable group structure with the lowest possible
social welfare--in this case, the configuration with groups of size
16 and 17. The following analysis shows how players wind up in this
suboptimal group structure.

\begin{figure}[h]
\includegraphics[scale=0.6]{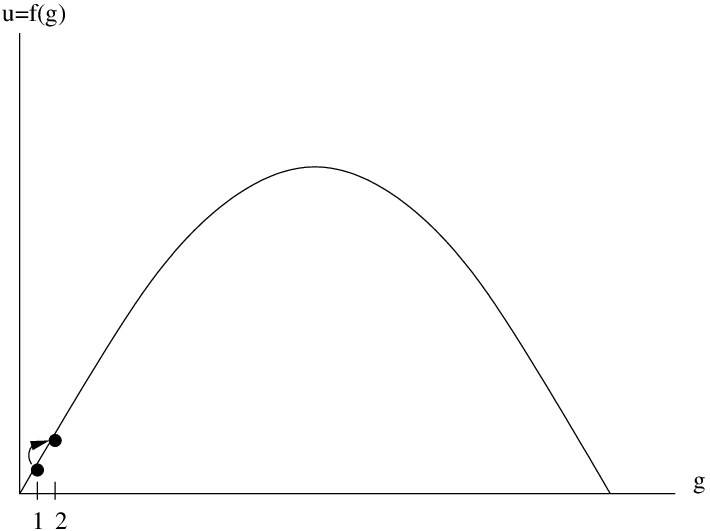}

\caption{\label{fig:One to Two} The first individual to move joins another
individual to form a group of size 2.}
\end{figure}

The players start the game as individuals, so the first player to
move faces a choice between remaining as an individual and forming
a group of size 2. She chooses the group of size 2 because it gives
her higher utility in the next period (Figure \ref{fig:One to Two}).

\begin{figure}[h]
\includegraphics[scale=0.6]{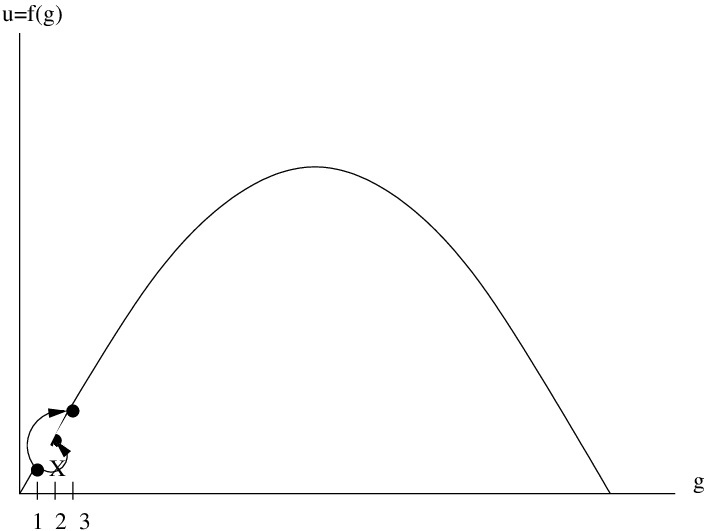}

\caption{\label{fig:One to Three}The second individual to move must choose
between forming a new group of size 2 or joining the existing group
of 3. She will choose the group of 3, since it gives her higher utility
than the group of 2.}
\end{figure}

The second player to move faces a similar choice--she must decide
whether to join the existing large group to form a group of 3, or
join another individual to form a second group of 2. The group of
3 gives her higher utility, so she joins that group (Figure \ref{fig:One to Three}).

\begin{figure}[h]
\includegraphics[scale=0.6]{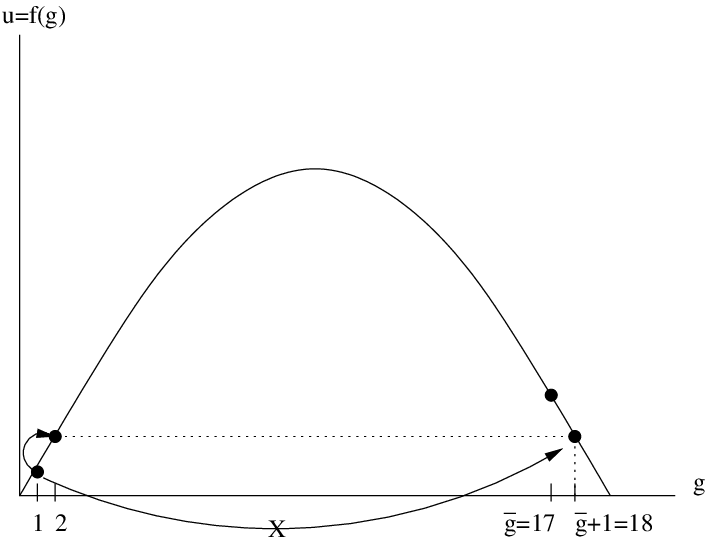}

\caption{\label{fig:new group forms} A new group forms when the large group
is size $\bar{g}=17$ because the individual is better off in a new
group of size 2. }
\end{figure}

A new group only forms when $f(2)\ge f(g+1)$ where $g$ is the size
of the existing large group. The smallest such $g$ is obviously $\bar{g}$,
in this case, a group of 17 (Figure \ref{fig:new group forms}). 

This is true regardless of how many {}``large'' groups (groups with
more than one individual) there are. Thus, the second group forms
when there are 83 individuals and one group of 17, the third group
forms when there are 69 individuals and two groups of size 17, and
so on. The last group forms when there are 15 individuals and five
groups of size 17.

This sixth group is the final group that will ever form. Individuals
may (and indeed, will) move between the existing groups, but no new
group will ever form, because no individual finds it advantageous
to move to a group of size 1. The individuals will stop moving when
all six groups are approximately the same size--namely, in the configuration
with two groups of size 16 and four groups of size 17: $\left\langle 16,16,17,17,17,17\right\rangle $.
As predicted by Theorem \ref{thm:number seq eq}, this is the stable
group arrangement with the lowest possible social welfare value. Note
also that at no point did we specify the order of play--thus, the
players will reach the arrangement $\left\langle 16,16,17,17,17,17\right\rangle $
regardless of their order of motion. 

Note that while it is tempting to ascribe this inefficient equilibrium
behavior to myopia, that is clearly not all that is at work. It is
possible to relax the myopia assumption, and still obtain the same
result.%
\footnote{Suppose individuals are forward-looking, but discount the future.
Then there exists a discount factor, $\delta\in\left(0,1\right)$,
such that forward-looking actors will reach the same, suboptimal equilibrium
outcome.%
} Moreover, myopia cannot why individuals would join a group that is
already too large. The key to the inefficiency is that when a new
member joins a group, her action alters the payoffs of every existing
member, creating an externality. When groups are smaller than $g^{*}$,
this externality is positive. However, when groups are larger than
$g^{*}$, the externality is negative--new members fail to internalize
the costs they impose on the incumbent membership. The result of this
externality is groups that are much larger than the socially optimal
size.

\section{Dynamic Group Formation with a Network Constraint}

The model of group formation from the previous section assumes that
individuals are free to join any existing group, regardless of its
current composition. In this section, I consider the effect of social,
spacial, and informational constraints on group membership decisions
and equilibrium behavior. I model these constraints as an exogenous
network of relationships between individuals. An individual cannot
join any group to which she is not connected on the network. I then
consider the relationship between the structure of the network constraint
and the efficiency of the equilibrium outcomes.

\subsection{Group Formation with a Network Constraint}

Consider a dynamic group formation game, but now, suppose individuals
have an exogenous network of connections to other people. An individual
can only join a group if it contains a person she is connected to
on the network. More formally, $\left(N,f(g),\phi,C\right)$ defines
a particular dynamic group formation game with a network constraint,
where $C$ is an exogenous, unchanging matrix of connections between
individuals--that is, $C_{ij}=1$ if $i$ is connected to $j$ on
the network and $0$ otherwise. In this game, an individual's action
set is restricted to include only those groups she is connected to:
$A_{i}(t)=\left\{ G\,|\, C_{ij}=1\, for\, some\, j\in G\right\} \cup\emptyset\subseteq\pi(t)\cup\emptyset$.
\footnote{Note that this differs significantly from the use of networks in \citet{Page2007a},
which uses a bipartite network to illustrate the partition of individuals
into groups--ie: each individual is linked to the group to which it
is a member. In this paper, the network links individuals to one another,
restricting an individual's choice of groups.%
}

Note that when the network is fully connected, every player knows
someone in every group and therefore $A_{i}(t)=\pi(t)\cup\emptyset$.
Thus, the unconstrained dynamic group formation game, above, is a
special case of the constrained game where the average degree is at
a maximum: $\left\langle d\right\rangle =N-1$. 

I will also find it convenient to refer to the connections between
groups, as well as the connections between individuals: for a network
constraint $C$, call two groups, $G_{j}$ and $G_{k}$ \emph{connected
}if $\exists h\in G_{j}$ and $i\in G_{k}$ such that $C_{hi}=1$.
That is, two groups are connected if there are individuals in the
two groups who are connected on the network, $C$.

\subsection{The Set of Stable Configurations with a Network Constraint}

\begin{figure}[h]
\includegraphics[scale=0.5]{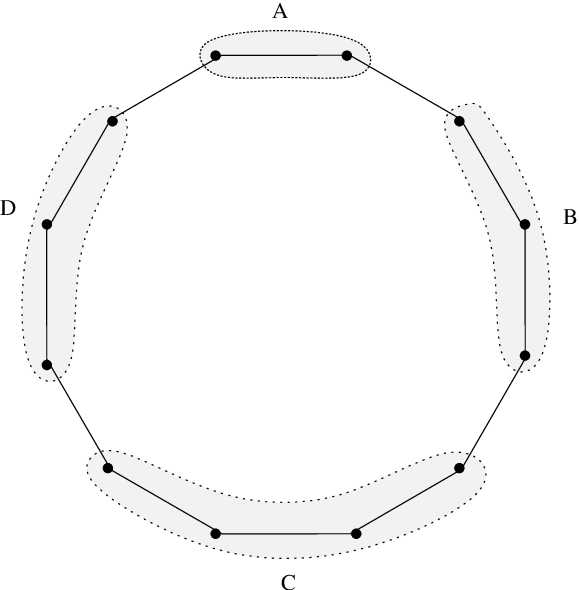}

\caption{\label{fig:Ring with new eq}An example of an equilibrium group configuration
on a ring. Note that this would not be an equilibrium on the fully-connected
network because the players in group C would move to group A. }
\end{figure}

As before, the equilibria of the dynamic group formation game must
be stable group configurations, which are the Nash equilibria of a
static game with the same constraint. An equilibrium of the static
game with a network constraint is a partition of the players into
groups that is both feasible and individually rational. In particular,
$\left(N,f(g),C\right)$ defines a static group formation game with
a network constraint. The set of Nash equilibria of this game, $\epsilon\left(N,f\left(g\right),C\right)$,
are the set of stable configurations of players into groups, $\left\{ G_{1}...G_{J}\right\} $,
such that $\forall i$, $i\in G_{j}$ implies: 
\begin{enumerate}
\item $f\left(g_{j}\right)\ge f\left(g_{k}+1\right)\forall G_{k}\in\left\{ G_{1}...G_{J}\right\} $ 
\item $C_{ij}=1$ for some $j\in G_{j}$ 
\end{enumerate}
Recall that in the unconstrained case, all stable group configurations
share two characteristics: groups are nearly all as large or larger
than the social optimum, and all groups larger than the social optimum
will be roughly the same size. One result of adding a network constraint
is that the there may exist stable group structures in which groups
are substantially different sizes.
\begin{claim}
For a given static group formation game $\left(N,f(g),C\right)$,
there may exist a Nash equilibrium group structure, $\left\{ G_{1}...G_{J}\right\} $,
such that $\left|g_{j}-g_{k}\right|>1$ for some $g_{j},g_{k}>g^{*}$.
\end{claim}
As an illustration of this claim, consider a game with 12 players
on a ring. Further suppose $g^{*}=2$ and $\bar{g}=6$, so that all
individuals want to be in a group of size 2, and will never form a
group larger than size 6. Figure \ref{fig:Ring with new eq} illustrates
a stable group structure of the static game $\left(N,f(g),C\right)$
with uneven group sizes. 

It is obvious from Figure \ref{fig:Ring with new eq} how the ring
affects the stability of this configuration. The individuals in group
C would like to join group A, but they are unable to because they
are not connected to that group on the social network. If the network
were fully connected, the individuals in group C would like to move
to group A, and the configuration would not be stable. Note that the
constraint of the ring could represent either a constraint on actions
(the players would move if they could) or information (the players
would move if they knew). It could also equally well represent an
explicit constraint (a legal constraint), an implicit constraint (a
social norm), or a functional constraint (a geographic coincidence).
This result provides some insight into an empirical puzzle. Our existing
models of group formation predict that groups will be the same size
in equilibrium. However, empirically, group sizes are clearly far
from identical. This model indicates that network-type constraints
may be one factor that leads to uneven group sizes. 

By exploiting the fact that any two connected individuals form a fully
connected subgraph, I can characterize all stable configurations in
a group formation game with a network constraint. 
\begin{thm}
\label{thm:Static Constrained Equilibria}Let $\left(N,f(g),C\right)$
be a static group formation game with single-peaked payoff function
$f(g)$ and network constraint $C$. $\left\{ G_{1}...G_{J}\right\} \in\varepsilon\left(N,f(g),C\right)$
if for all connected groups, $G_{j}$ and $G_{k}$, either \end{thm}
\begin{enumerate}
\item $g_{j},\, g_{k}\ge g^{*}$ and $|g_{j}-g_{k}|\le1$

or

\item $g_{j}<g^{*}\le g_{k}$ and $f(g_{k})\ge f(g_{j}+1)\ge f(g_{j})\ge f(g_{k}+1)$ \end{enumerate}
\begin{proof}
Simply note that any pair of connected groups contains a pair of connected
agents, who form a fully connected subgraph of the original graph.
The result above follows immediately from Theorem \ref{thm:static eq}. 
\end{proof}

\subsection{\label{sub:Seq Constrained}Equilibrium Behavior}

When the players in a group formation game face a binding network
constraint,%
\footnote{That is, one that actually restricts their action sets.%
} their equilibrium behavior is much more complicated than it is in
the unconstrained case. A simple example (see Appendix) shows that
Theorem \ref{thm:number seq eq} need not hold when there is a network
constraint--the set of equilibria may depend on the order of play
and there will often be multiple equilibrium group size configurations. 

\begin{figure}[h]
\includegraphics[scale=0.5]{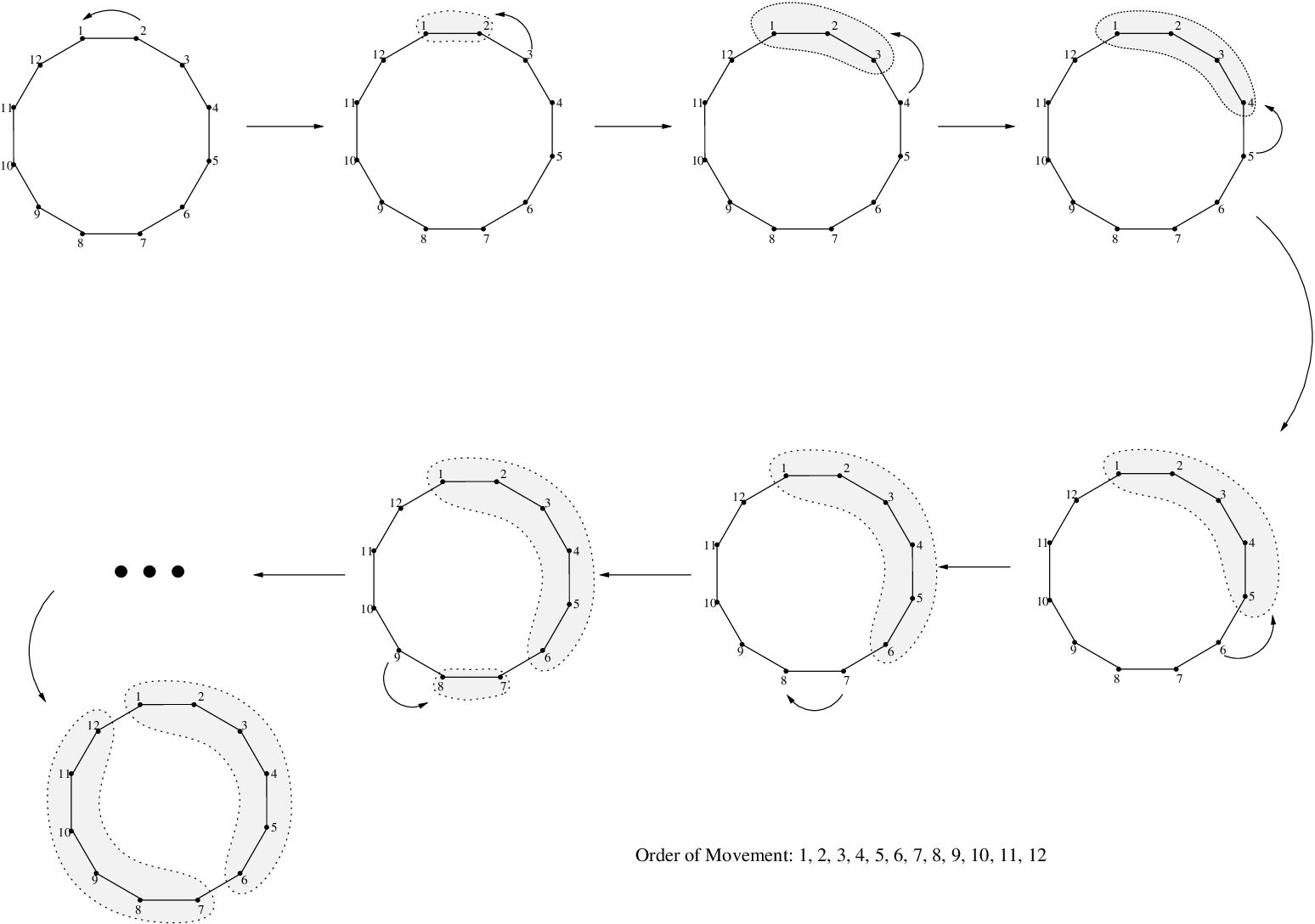}

\caption{\label{fig:Seq in Order Around Ring} In this game, 12 individuals
are arranged in a ring. The payoff function, $f(g)$, has maximum
$g^{*}=2$ and $\bar{g}=6$. The individuals move in order around
the ring--$\phi_{1}=1,2,3,4,5,6,7,8,9,10,11,12$--and wind up in two
groups of size 6. In fact, $\left\langle 6,6\right\rangle $ is the
only equilibrium group size configuration of the game $\left(12,f(g),\phi_{1}\right)$.
Figure \ref{fig:Seq Groups of 2} shows the same game with a different
order of play.}
\end{figure}

Consider a game with 12 players arranged in a ring, with a single-peaked
payoff function, $f(g)$ such that $g^{*}=2$ and $\bar{g}=6$. Now,
consider two different orders of motion: $\phi_{1}$ and $\phi_{2}$.
For the first order of motion, suppose that the players proceed in
order around the ring--that is, $\phi_{1}=$ $\left(1,2,3,4,5,6,7,8,9,10,11,12\right)$.
Figure \ref{fig:Seq in Order Around Ring} shows game play leading
to an equilibrium coalition structure with two groups of size 6. A
simple analysis indicates that this is the only equilibrium configuration
for this order of motion. %
\footnote{Because of the order of play, the individuals are always choosing
between joining an existing large group, forming a new group of two,
or remaining as an individual. This choice is the same as the choice
players face in the unconstrained game with the same payoff function,
and therefore, the players must reach the same equilibrium coalition
structure as they do in the unconstrained game: $\left\langle 6,6\right\rangle $.%
} 

\begin{figure}[h]
\includegraphics[scale=0.5]{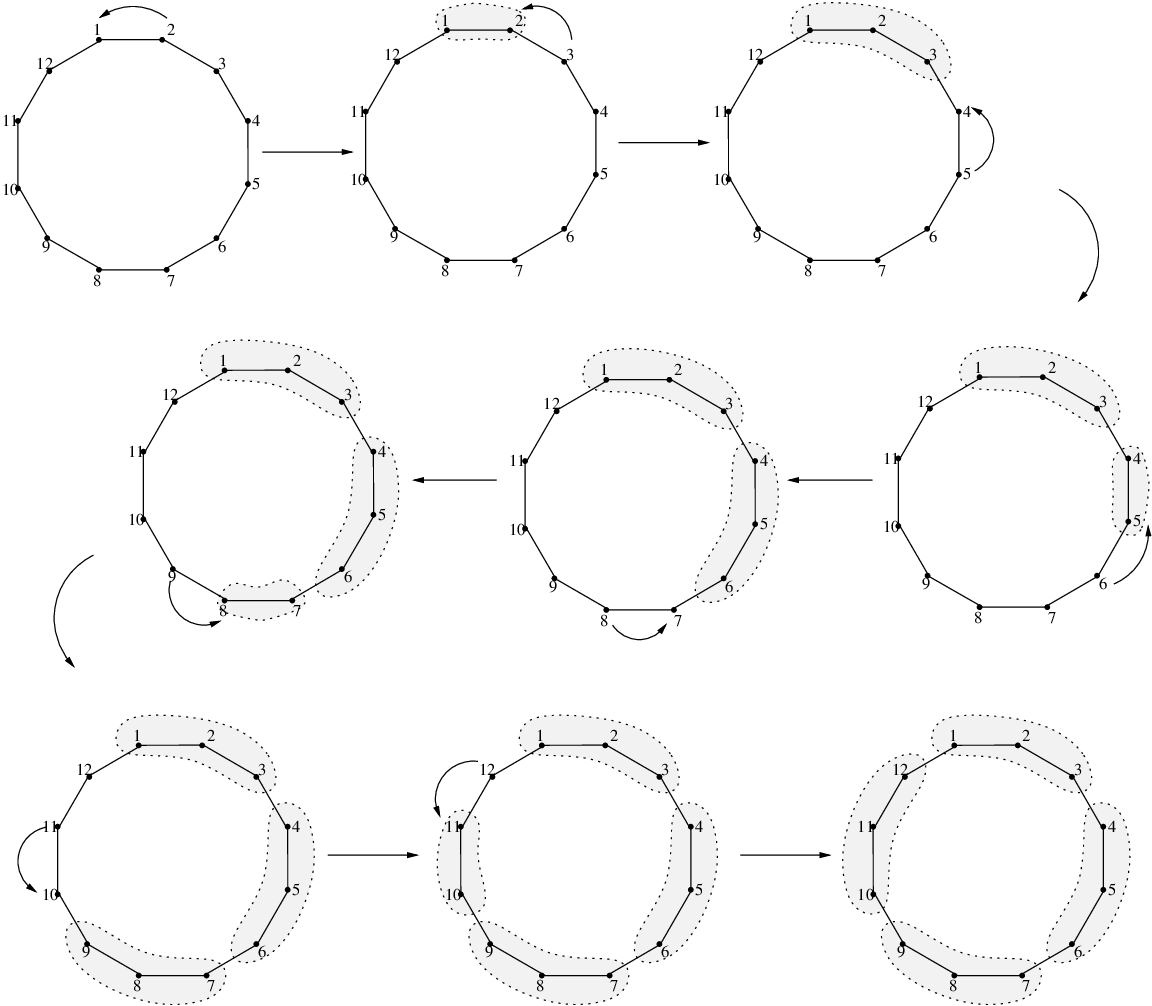}

\caption{\label{fig:Seq Groups of 2}This game is identical to the game presented
in Figure \ref{fig:Seq in Order Around Ring} except for the order
of motion: $\phi_{2}=2,3,5,6,8,9,11,12,1,7,4,10$. This figure shows
a particular sequence of moves, which leads to groups of the ideal
size: $\left\langle 3,3,3,3\right\rangle $. Note that $\left\langle 3,3,3,3\right\rangle $
is not an equilibrium of the game presented in Figure \ref{fig:Seq in Order Around Ring},
proving that the set of equilibria may depend on the order of play.}
\end{figure}

Now consider a second game with the same number of players, network
constraint, and payoff function, but a different order of play $\phi_{2}=\left(2,3,5,6,8,9,11,12,1,7,4,10\right)$.
Figure \ref{fig:Seq Groups of 2} shows one possible sequence of game
play, given $\phi_{2}$. 
Because the first few players to move are separated from the existing
large groups, they are unable to impose on the groups that have already
formed, as they did in the previous example. The result is an equilibrium
coalition structure with four groups of the ideal size: $\left\langle 3,3,3,3\right\rangle $.
Since $\left\langle 3,3,3,3\right\rangle $ is in $\varepsilon\left(N,f(g),\phi_{2},C\right)$
but not in $\varepsilon\left(N,f(g),\phi_{1},C\right)$, it is clear
that the order of motion does affect the set of equilibria.

\begin{figure}[h]
\includegraphics[scale=0.5]{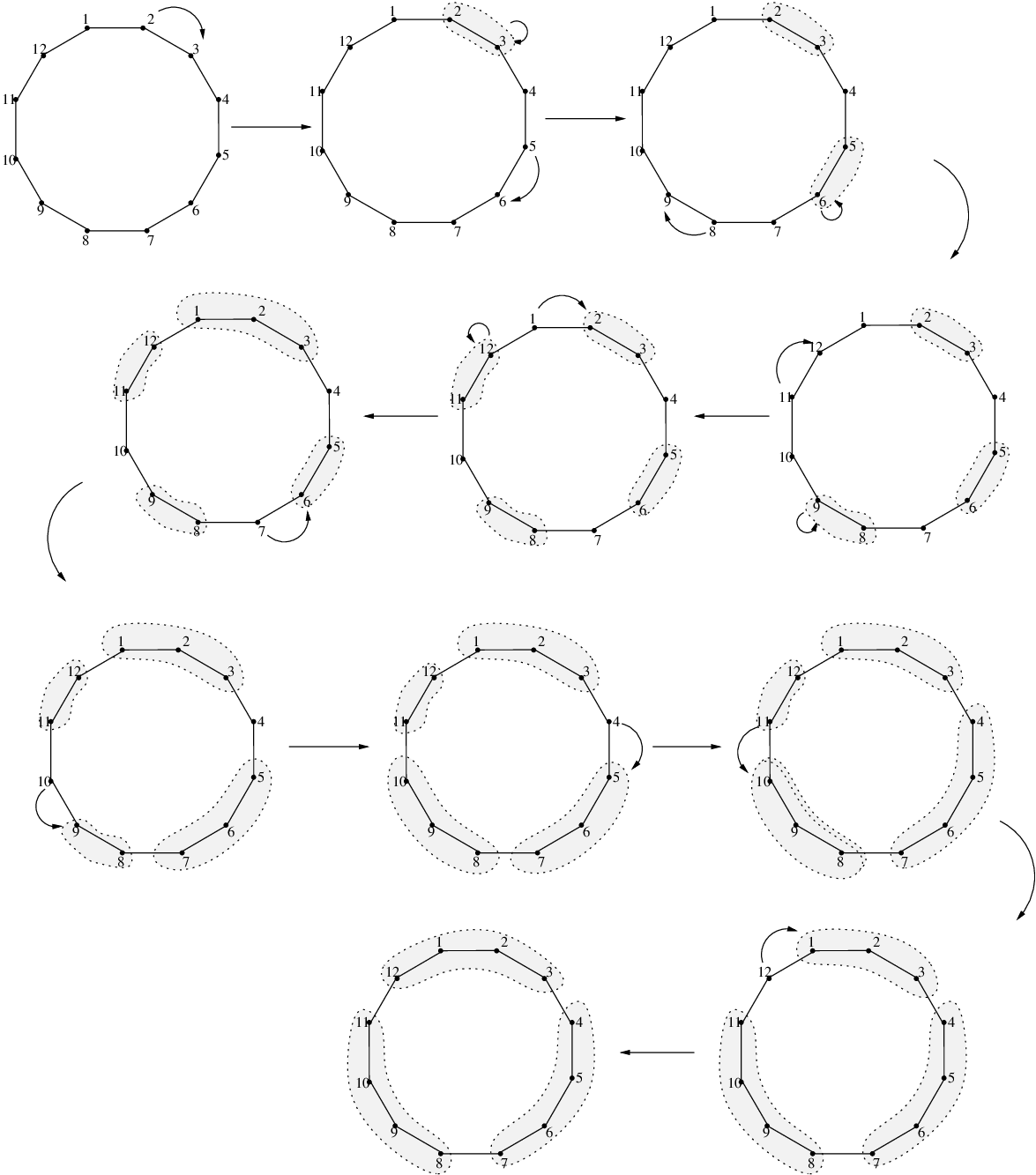}

\caption{\label{fig:Seq Groups of 3} This game is identical to that presented
in Figure \ref{fig:Seq Groups of 2}. Note, in particular, that the
order of play is the same: $\phi_{2}=2,3,5,6,8,9,11,12,1,7,4,10$.
However, the players have made different random choices, leading to
a different equilibrium outcome: $\left\langle 4,4,4\right\rangle $.
This shows that when players are sufficiently constrained, there need
not be a unique equilibrium coalition size configuration.}
\end{figure}

Of course, the outcome pictured in Figure \ref{fig:Seq Groups of 2}
is not the only possible equilibrium of the game with order of play
$\phi_{2}$. Many players in this game are forced to make random choices.
Figure \ref{fig:Seq Groups of 3} shows that if some of those players
make different choices, then the players will find themselves in a
different configuration--in this case, $\left\langle 4,4,4\right\rangle $.

Note that when players move according to $\phi_{2}$, the equilibrium
behavior produces groups closer to the socially optimal size, and
thus a higher social welfare than the equilibrium outcome in the unconstrained
game. Thus, imposing the network constraint actually improves equilibrium
outcomes.

\subsection{Network Structure and Efficiency}

Given that equilibrium behavior on a complete network (eg: the unconstrained
case) is different than the equilibrium behavior on the ring, one
natural question is how the structure of the network constraint will
affect the efficiency of the resulting equilibrium, where by efficiency,
I mean the fraction of the maximum social welfare captured by the
players: $\omega\left(\left\langle g_{1}...g_{J}\right\rangle \right)=\left(\frac{\sum_{j\in\left\langle g_{1}...g_{J}\right\rangle }g_{j}f\left(g_{j}\right)}{Nf\left(g^{*}\right)}\right)$.
In particular, I will consider the efficiency of equilibria on a well-studied
class of network structures, called Watts-Strogatz networks.%
\footnote{\citet{Watts1998}%
} 

A Watts-Strogatz network is designed to model a wide range of different
types of network structures, using only two parameters. It is constructed
as follows. One starts with a regular network of degree $d$--this
is a network in which every individual is connected to her $\frac{d}{2}$
nearest neighbors on each side. Then each of the links in the regular
network is rewired with probability $p$. A link is rewired by disconnecting
one end and reconnecting it to a different, random node in the network. 

\begin{figure}[h]
\includegraphics[scale=0.5]{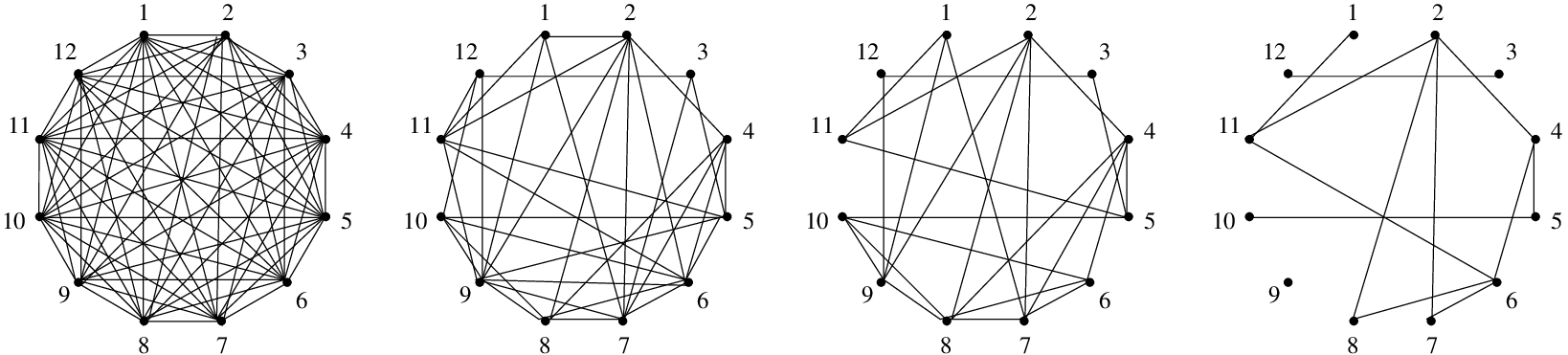}

\caption{\label{fig:Degree Graphs}Four different networks with different average
degree $\left(d\right)$. Networks with lower degree represent a more
binding constraint on group membership decisions. }
\end{figure}

\begin{figure}[h]
\includegraphics[scale=0.5]{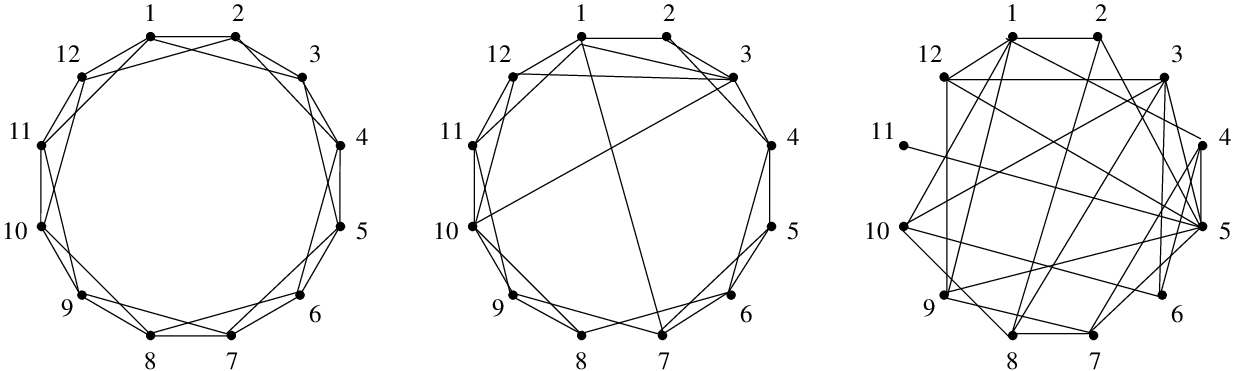}

\caption{\label{fig:Watts Strogatz networks} Three networks with different
Watts-Strogatz parameters $\left(p\right)$. The Watts-Strogatz network
starts with a regular network, where nodes are connected to their
two nearest neighbors on each side. Each of these links is rewired
at random with probability $p$. Thus, as $p$ increases, the network
becomes increasingly random. }
\end{figure}

The structure of the Watts-Strogatz network is controlled by adjusting
these two parameters: $d$ and $p$. The first, $d$, is the density
of the network--the average number of links per person--and it reflects
how binding the constraint is (see Figure\ref{fig:Degree Graphs}).
On one extreme is a network where everyone is connected to everyone
else--where $d=N-1$. As mentioned above, individuals on this network
can join any group in the system, so a game on this network is equivalent
to the unconstrained case. On the other extreme is a network where
no one is connected to anyone else--where $d=0$. In this case, the
network is completely disconnected, and nobody can join any group.
This is the network constraint at its most binding. The second parameter
in the network is the Watts-Strogatz parameter, $p\in\left(0,1\right)$.
This parameter indicates what fraction of the links are made at random,
and what fraction remain regular (see Figure \ref{fig:Watts Strogatz networks}).
Adjusting this parameter allows me to examine a spectrum of different
network types--when $p=0$, the network is regular and approximates
a spatial network; when $p=1$, the individuals are connected at random;
for values of $p$ between 0 and 1, the network has a {}``small world''
structure, which approximates that of a social network. A pair $\left(d,p\right)$
describes a family of networks with similar topological characteristics.
 
Because of the multiplicity of equilibria for a single group formation
game, it is necessary to look at the average outcome over a large
number of games played on networks with the same values of $d$ and
$p$.%
\footnote{An alternative method would be to determine the distribution of outcomes
combinatorially and calculate the expected social welfare exactly.
However, this method would yield results that are overly narrow, applying
only to the specific network considered. As discussed early, I would
like to draw conclusions about a {}``class'' of networks with similar
topologies, which is why I choose to average over a large number of
games played on topologically similar, but not identical, networks.%
} Players play 1000 group formation games on social networks with the
same parameters, $\left(d,p\right)$. I report how close the equilibrium
group size vector, $\left\langle g_{1}...g_{J}\right\rangle $, is
to being optimal. In particular, I report the fraction of the maximum
possible social welfare that the players realize, in equilibrium:
\[
\omega\left(\left\langle g_{1}...g_{J}\right\rangle \right)=\left(\frac{\sum_{j\in\left\langle g_{1}...g_{J}\right\rangle }g_{j}f\left(g_{j}\right)}{Nf\left(g^{*}\right)}\right)
\]
$\omega=1$ indicates that the players all found themselves in groups
of the idea size. Lower values of $\omega$ indicate greater inefficiency.

Here, I present results for a game with logistic payoffs and $g^{*}=10$.
The results are the same for other single-peaked payoff functions. 

\begin{figure}
\includegraphics[width=0.9\columnwidth]{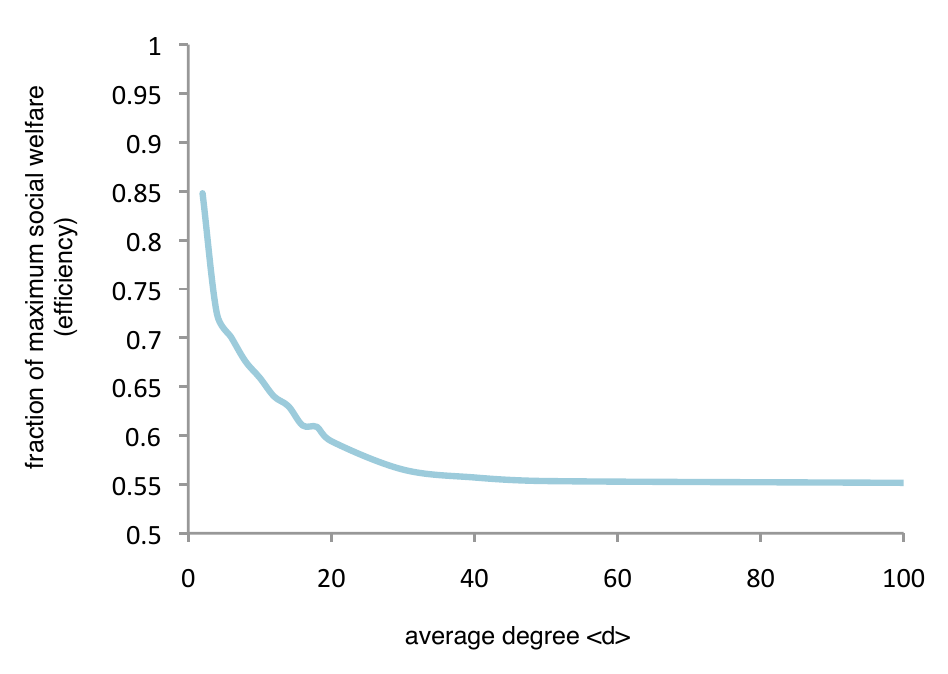}

\caption{\label{fig:Efficiency by Degree}Efficiency, as a function of the
degree of the network constraint. Individuals form more efficiently-sized
groups when constrained by a network of lower degree. }
\end{figure}

Figure \ref{fig:Efficiency by Degree} illustrates the relationship
between the density of the underlying network and the efficiency of
the resulting group structure for a random network $\left(p=1\right)$.
Since the size of an individual's action set is bounded above by her
degree on the network, degree reflects how constraining the network
is on individual behavior. As the degree of the network decreases,
the players extract a greater fraction of the maximum possible social
welfare. The resulting group structure is more efficient than when
the players are unconstrained. 

\begin{figure}
\includegraphics[width=0.9\columnwidth]{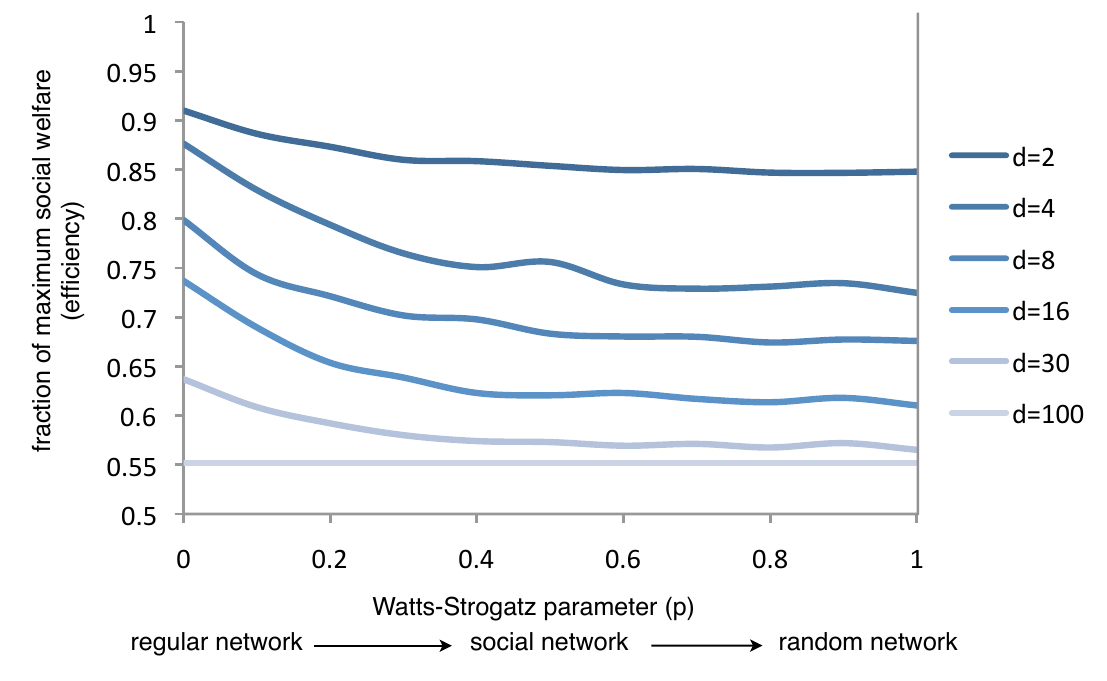}

\caption{\label{fig:Efficiency by WS}Efficiency, as a function of the Watts-Strogatz
parameter. The Watts-Strogatz parameter is a measure the amount of
order in the network. The group structure is less efficient the more
random the underlying network constraint is.}
\end{figure}

Figure \ref{fig:Efficiency by WS} shows the effects of the Watts-Strogatz
parameter on social welfare. Recall that the Watts-Strogatz parameter,
$p$, is the fraction of the links in the network that are made at
random. For networks of all degree, the group structure becomes less
efficient as the underlying network becomes more random. This is because
the more random connections there are, the less binding the network
constraint will be for the average individual. When only a small fraction
of connections are random, the average individual will tend to know
lots of people in the same group. When connections are random, they
will know individuals in many different groups, and the constraint
will have less bite. 

\section{\label{sec:Discussion-and-Conclusion}Discussion and Conclusion}

The equilibrium of the dynamic group formation game is interesting
because it is so clearly inefficient--groups become much larger than
is socially optimal, even when all individuals agree on the optimum
size. Groups become too large because of an externality that new group
members impose on existing group members. When a new member joins
a group, she alters the utility of all existing group members, creating
an externality. When the group is smaller than the social optimum,
that externality is positive. However, when the group is the optimal
size, the externality is a negative one. The entering member is obviously
made better off by joining (otherwise, she would not join), but the
rest of the group is made worse off. The negative externality causes
individuals to enter a group that does not benefit from the extra
member, which then drives groups to become too large. One way to interpret
this result is that starting a new group is more difficult than joining
an existing, larger group. Thus, there is an incentive to free ride
off of early group founders. New groups are under-provided, while
existing groups become too large. New groups only form when existing
groups become so large that the relatively lower payoffs from starting
a new group become worthwhile. Only at that point will groups splinter. 

I have shown that the social, spacial, and informational constraints
faced by individuals making group membership decisions will improve
the efficiency of the resulting group structure. Initially, it might
be surprising that restricting individuals by forcing them to choose
groups that they have a connection to would improve outcomes. However,
this is consistent with the fact that the inefficiency in the unconstrained
case is due to a negative externality. When players are constrained
by social, spacial, and informational networks, they do not always
have access to established groups and are forced to form new groups,
rather than joining groups that are already too large. In other words,
the network constraint limits the players' choice sets, which forces
them to internalize the start up costs of creating a new group, limiting
their ability to impose a negative externality on others. 

There is a clear relationship between the structure of the network
constraint and the equilibrium group structure. The players form into
more efficient groups when they are constrained by networks that have
lower density (lower average degree, $d$) and are more ordered and
less random (smaller fraction of random links, $p$). Both of these
aspects of network topology affect how binding the network constraint
is. A network with lower average degree means that individuals on
the network will be connected to fewer groups on average, restricting
their action set and making the network constraint more binding. Similarly,
when there are few random connections in the network, the average
individual will have access to fewer distinct groups, making the network
constraint bind. This suggests that group structures will be more
efficient when group membership requires a stronger relationship (implying
a network of lower density). Moreover, given that social networks
have more random links than spacial networks, we would expect groups
to be closer to the ideal size when group membership is constrained
by geography, rather than social connections.

\section{Appendix: }

\subsection{Proof of the Static Game Equilibrium}

Theorem \ref{thm:static eq} is built up from three lemmas. The first
lemma states that in any static equilibrium, at most one group will
be smaller than the socially optimal size. The second lemma states
that all groups larger than the optimum will be the same size, up
to integer constraints. The third lemma pins down the size of any
group smaller than the optimum. 
\begin{lem}
\label{lem:at most one small}Let $\left(N,f(g)\right)$ be a static
group formation game with $f(g)$ single-peaked. Then $\exists$ no
equilibrium $\left\langle g_{1}...g_{J}\right\rangle \in\varepsilon\left(N,f(g)\right)$
such that $g_{i}\le g_{j}<g^{*},\, i\ne j$. That is, in equilibrium
at most one group will be smaller than the social optimum. \end{lem}
\begin{proof}
Towards a contradiction, suppose $\exists\left\langle g_{1}...g_{J}\right\rangle \in\varepsilon\left(N,f(g)\right)$
such that $g_{1}\le g_{2}<g^{*}$. $f(.)$ is strictly increasing
in that range, so $f(g_{1})<f(g_{2}+1)$. But then players in group
1 have an incentive to move to group 2, so $\left\langle g_{1}...g_{J}\right\rangle $
cannot be an equilibrium 
\end{proof}
Lemma \ref{lem:at most one small} implies that in characterizing
$\varepsilon\left(N,f(g)\right)$, we need consider only two cases:
either all of the groups are larger than the socially optimal size
($g_{1}...g_{k}\ge g$), or exactly one group is small ($g_{1}<g^{*}$and
$g_{2}...g_{k}\ge g^{*}$). The following two Lemmas address the sizes
of the groups in these two different cases. Lemma \ref{lem:all same size}
shows that in any equilibrium where all groups are larger than the
social optimum, the groups must be approximately the same size. Lemma
\ref{lem:-odd size} sets a more restrictive condition in the case
where one group is smaller than the social optimum.
\begin{lem}
\label{lem:all same size}Let $\left(N,f(g)\right)$ be a static group
formation game with $f(g)$ single-peaked. Then for all $\left\langle g_{1}...g_{J}\right\rangle \in\varepsilon\left(N,f(g)\right)$,
$|g_{i}-g_{j}|\le1\,$$\forall g_{i},g_{j}\ge g^{*}$. That is, in
equilibrium, all groups larger than the social optimum must be the
same size, up to integer constraints. \end{lem}
\begin{proof}
Towards a contradiction, suppose $\exists\left\langle g_{1}...g_{J}\right\rangle \in\varepsilon\left(N,f(g)\right)$
such that $g_{1}>g_{2}\ge g^{*}$ and $g_{1}-g_{2}>1$. $f(.)$ is
strictly decreasing in this range, so $f(g_{1})<f(g_{2}+1)$ But then
players in group 1 have an incentive to move to group 2, so $\left\langle g_{1}...g_{J}\right\rangle $
cannot be an equilibrium. 

Note that this result extends a result in Nitzen (1991) to the case
of single-peaked utility. Arnold and Wooders (2005) prove a similar
result for a sequential game. The following lemma extends that result
to the case where one group is smaller than the social optimum. The
Nash Equilibrium requires a slightly stronger restriction on the size
of the groups.\end{proof}
\begin{lem}
\label{lem:-odd size} Let $\left(N,f(g)\right)$ be a static group
formation game with $f(g)$ single-peaked. Then for all $\left\langle g_{1}...g_{J}\right\rangle \in\varepsilon\left(N,f(g)\right)$
such that $g_{1}<g^{*}$%
\footnote{By Lemma \ref{lem:at most one small}, this implies $g_{l}\ge g^{*}\forall l\ne k$%
}, both of the following must be true: \end{lem}
\begin{enumerate}
\item $f(g_{j})\ge f(g_{1}+1)\ge f(g_{1})\ge f(g_{j}+1)\,\forall\, j>1$ 
\item $g_{j}=g_{k}\,\forall\, j,k\ne1$ \end{enumerate}
\begin{proof}
Let $\left\langle g_{1}...g_{J}\right\rangle \in\varepsilon\left(N,f(g)\right)$such
that $g_{1}<g^{*}$.

Part 1: Consider group 1 (the small coalition) and an arbitrary group
$j$, such that $g_{j}\ge g^{*}$ Note that $f(g_{1})<f(g_{1}+1)$
and $f(g_{j})>f(g_{j}+1)$. If $f(g_{1})<f(g_{j}+1)$, then players
in group 1 would move to group $k$. Similarly, if $f(g_{j})<f(g_{1}+1)$,
then players in group $k$ would move to group 1. Together, these
three inequalities imply $f(g_{j})\ge f(g_{1}+1)\ge f(g_{1})\ge f(g_{j}+1)\,\forall\, j>1$

Part 2: consider two arbitrary groups, $j$ and $k$, such that $g_{k}\ge g_{j}\ge g^{*}$.
Lemma \ref{lem:all same size} indicates that $g_{k}-g_{j}\le1$.
Towards a contradiction, suppose $g_{k}-g_{j}=1$, so that $g_{k}=g_{j}+1$.
By Part 1, $f(g_{j}+1)\le f(g_{1})$. Since we assumed $g_{k}=g_{j}+1$,
this implies that $f(g_{k})\le f(g_{1}).$ But since $f(g_{1})<f(g_{1}+1)$
to the left of the optimum, $f(g_{k})<f(g_{1}+1)$, meaning that players
in group $j$ would move to group $k$. Thus, it must be that $g_{j}=g_{k}$
exactly.
\end{proof}
Together, the restrictions imposed by these three lemmas form the
basis of Theorem \ref{thm:static eq}.

\subsection{Proof of multiple stable configurations}

Theorem \ref{thm:number of static eq} puts a lower bound on the number
of equilibria in the set $\varepsilon\left(N,f(g)\right)$, showing
that there the static game has multiple equilibria, meaning that there
will be multiple stable group configurations.
\begin{thm}
\label{thm:number of static eq}Let $\left(N,f(g)\right)$ be a static
group formation game. Then $\left|\varepsilon\left(N,f(g)\right)\right|\ge\frac{N}{g^{*}}-\frac{N}{\bar{g}}-1$. \end{thm}
\begin{proof}
I will set the lower bound by enumerating the equilibria in which
all groups are larger than the social optimum (ie: the first set in
Theorem \ref{thm:static eq}). Note that since all groups are approximately
the same size, each equilibrium with all large groups is entirely
characterized by the \emph{number} of groups. The largest possible
group is $\bar{g}$ and the smallest possible group is $g^{*}$. Thus,
there should be one equilibrium for each integer in the interval $\left[\frac{N}{\bar{g}},\frac{N}{g^{*}}\right]$,%
\footnote{This is actually also a lower bound on the number of equilibria with
all large groups. There could be more, depending on whether $g^{*}$
and $\bar{g}$ divide $N$ evenly, but including that complication
only adds more equilibria, keeping the lower bound accurate (albeit
a bit lower than is strictly necessary).%
} or $\frac{N}{g^{*}}-\frac{N}{\bar{g}}-1$. 
\end{proof}
Since the lower bound in Theorem \ref{thm:number of static eq} is
usually greater than 1, the static game will usually have multiple
equilibria.

\section{Acknowledgements}

Thanks to Scott Page, Rick Riolo, Robert Willis, Lada Adamic, and
Ross O'Connell. This work was done with the support of the NSF. Computing
resources provided by the University of Michigan Center for the Study
of Complex Systems.

\bibliographystyle{elsarticle-harv}
\bibliography{library}

\end{document}